\documentstyle[12pt]{article}

\newcommand{\F}{\noindent}

\newcommand{\SP}{\smallskip}
\newcommand{\MP}{\medskip}
\newcommand{\BP}{\bigskip}

\newcommand{\beq}{\begin{eqnarray}}
\newcommand{\ene}{\end{eqnarray}}
\newcommand{\beqn}{\begin{eqnarray*}}
\newcommand{\enen}{\end{eqnarray*}}

\newcommand{\eq}[1]{(\ref{#1})}

\begin{document}

\ 

%\rightline{KIMS-2001-02-11}
%\rightline{gr-qc/0102057}
\BP

\vskip12pt

\vskip8pt

\begin{center}

\Large

{\bf Quantum Mechanical Clock and 
\MP

Classical Relativistic Clock}
\normalsize

\BP

\vskip18pt

\author{Hitoshi Kitada}
\SP

%\address{
%Department of Mathematical Sciences,
%University of Tokyo,
%Komaba, Meguro, Tokyo 153-8914, Japan, \\
%e-mail: kitada@ms.u-tokyo.ac.jp,
%http://kims.ms.u-tokyo.ac.jp/}

%\maketitle

Department of Mathematical Sciences

University of Tokyo

Komaba, Meguro, Tokyo 153-8914, Japan

e-mail: kitada@ms.u-tokyo.ac.jp

http://kims.ms.u-tokyo.ac.jp/

\end{center}

\vskip8pt

\begin{abstract}
\F
Quantum mechanical clock is introduced as a means measuring the 
parameter of quantum mechanical motion, and is seen
 to be equivalent to the relativistic classical clock.
\end{abstract}

\SP

\vskip18pt

\F
{\bf Introduction}

\BP

As an introductory question, let us consider the light clock which Einstein used in his definition of an ideal clock in special theory of relativity. A light clock consists of two mirrors stood parallel to each other with light running mirrored to each other continuously. The time is then measured as the number of counts that the light hits the mirrors. This clock is placed stationary to an inertial frame of reference, and the time of the frame is defined by the number of the light-hits of this clock. Insofar as the light is considered as a classical wave and the frame is an inertial one which has no acceleration, this clock can measure the time of the frame. One feature of this clock is that the time is defined by utilizing the distance between the two mirrors and the velocity of light in vacuum which is assumed as an absolute constant in special theory of relativity. Thus time is measured only after the distance between the mirrors and the velocity of light are given, and it is not that time measures the motion or velocity of light wave. 

This light clock occupies a certain volume in space, and if an acceleration exists, the mirrors in the clock should be distorted according to general theory of relativity. In this case the number of counts of light between two mirrors cannot be regarded as giving the time of a certain definite frame of reference. To any point in the clock, different metrical tensor may be associated, and one cannot determine the time of which point the clock measures. Here appears a problem of the size of the actual clock which cannot be infinitesimally small. In this sense, the operational definition of the clock in general theory of relativity has a problem. It seems that this problem may be avoided by an interpretation that the general theory of relativity is an approximation of reality, and the theory gives a sufficiently good approximation as experiments and astronomical observations show. This problem, however, will be a cause of difficulty when one tries to quantize the field equation of general theory of relativity, for the expected quantized theory should be covariant under the diffeomorphism which transforms a point of a space-time manifold to a point of the manifold, and no point can accommodate any clocks with actual sizes.

To have sound foundations to resolve these problems, we therefore have to find, firstly, a definition of clock and time which should be given through length (or positions) and velocity (or motion) to accord with the spirit of Einstein's light clock, and secondly, our notion of time should have a certain ``good" residence just as the inertial frame of reference in special theory of relativity accommodates the light clock. 

As a residence, I prepare a Euclidean quantum space, and within that space I define a quantum-mechanical clock which measures the common parameter of quantum-mechanical motions of particles in a (local) system consisting of a finite number of particles. Since clocks thus defined are proper to each local system, and local systems are mutually independent as concerns the relation among the coordinates of these systems, we can impose relativistic change of coordinates among them. And the change of coordinates gives a relation among those local systems which would yield relativistic quantum-mechanical Hamiltonians, explaining the actual observations.

These are an outline of what I did in Refs. \cite{Ki} and \cite{Ki2}. The problem between quantum mechanics and relativistic theories has been noted \cite{B} already in 1932 by John von Neumann in the footnote on page 6 of the English translation \cite{N} of his book ``Die Mathematische Grundlagen der Quantenmechanik," Springer-Verlag, Berlin:
\begin{quotation}
\F
-- in all attempts to develop a general and relativistic theory of
electromagnetism, in spite of noteworthy partial successes, the theory (of
Quantum Mechanics) seems to lead to great difficulties, which apparently
cannot be overcome without the introduction of wholly new ideas.
\end{quotation}

I will consider the problem in the case of quantum mechanics and special theory of relativity, and give a solution below in the case of one particle in 3 dimensional space $R^3$ first. After then I will add a sketchy explanation of general $N$-particle case, which is almost obvious by the first part.
\BP

\F
{\bf I. One particle case with mass $m$}
\BP

\F
We consider this one particle in the universe expressed as a vector bundle $X\times R^6$ with base space $X$ being the curved Riemannian space where relativistic CM (classical mechanics) holds and with a fibre $R^6$ (a phase space associated to each point $x\in X$) with an Euclidean structure on which QM (quantum mechanics) holds as in Refs. \cite{Ki}, \cite{Ki2}. Then inside the QM system of the one particle (which we call a ``local system" of the particle), the particle follows the Schr\"odinger equation when the local time $t$ of the system is given as in definition 1 below, and in the classical space outside the local system the particle is observed as a classical particle moving with velocity $v$ relative to the observer in the observer's time.
\MP

\F
The motion inside a local system corresponds to the usually
 conjectured ``invisible" {\it zitterbewegung} of particles like
 virtual photons associated around elementary particles. This
 motion exists even when the particle is at ``zero-point"
 energy, e.g. Not only that but the spin, the vortices observed
 at very low temperatures in superfluidity etc., would be
 interpreted \cite{Natarajan} as this internal motion inside
 a local system.
\MP

\F
Instead of imposing connections among those fibres
 as is usually expected for the terminology 
``vector bundle," we postulate a relation between
 the internal and external worlds by actualizing the
 ``invisible" internal motion as velocity $u$ inside a
 local system in the following two axioms, which is
 assumed to hold between the two worlds on the
 occasion of observation:
\MP

\F
A1. Let $u$ be the QM velocity of the particle inside the local system. Then $u$ and
$v$ satisfy
$$
|u|^2+|v|^2=c^2,
$$
where $c$ is the velocity of light in vacuum.
\MP

\F
A2. The magnitude of momentum inside the local system is observed constant
independent of the velocity $v$ relative to the observer:
$$
m^2|u|^2=(m_0)^2c^2,
$$
where $m_0$ is the rest mass of the particle and $m$ is the observed mass of the
particle moving with velocity $v$ relative to the observer.
\BP

\F
These axioms are an extension of Einstein's principle \cite{E} of the constancy of the velocity of light in vacuum to a principle of the constancy of the velocity of everything when the velocities in the internal quantum mechanical space and the external classical relativistic space are summed. See Natarajan \cite{Na} for the natural motivation for these axioms corresponding to postulates IV and V in Ref. \cite{Na}, whereas Natarajan considers both of internal and external worlds are classical. We consider the internal world quantum mechanical. Thus the above axioms need a justification in order for these to have a consistent meaning, which discussion is given below. This will then yield a unification of quantum mechanics and special theory of relativity.

\BP

\F
To see the consequences of those axioms, we first consider the internal motion inside the QM local system of the particle.
\MP

\F
Let $H$ be the Hamiltonian of the particle inside the QM space:
\beq
H = -(\hbar^2)/(2m)(\Delta_x)
= \frac{1}{2m}\left(\frac{\hbar}{i}\right)^2\left(\frac{\partial}{\partial x}\right)^2 = \frac{1}{2m}P^2, \label{(1)}
\ene
where
$$
\frac{\partial}{\partial x}=\left(\frac{\partial}{\partial x_1},\frac{\partial}{\partial x_2},\frac{\partial}{\partial x_3}\right),
\quad
\left(\frac{\partial}{\partial x}\right)^2=\left(\frac{\partial}{\partial x_1}\right)^2+\left(\frac{\partial}{\partial x_2}\right)^2+\left(\frac{\partial}{\partial x_3}\right)^2,
\quad
P=\frac{\hbar}{i}\frac{\partial}{\partial x}
$$
are partial differentiation with respect to 3 dimensional space coordinate
$x=(x_1,x_2,x_3)$ and $\hbar=h/(2\pi)$ with $h$ being Planck constant. $P$ is identified with the QM momentum of the particle.
\MP

\F
We now define a clock of the local system. A best definition of a clock of a local system would be the evolution of the local system itself: $\exp(-itH/\hbar)$, where $H$ is the quantum mechanical Hamiltonian of the system. Then the local time of the system is defined by the $t$ on the exponent of the clock $\exp(-itH/\hbar)$:

\MP

\F
Definition 1.
We call $\exp(-itH/\hbar)$ a clock of the local system of the one particle,
and the $t$ in $\exp(-itH/\hbar)$ the local time of the system.
\MP

\F
Then we can show that this $t$ equals the time defined by using a clock of any part of the system up to the error that is consistent with the uncertainty principle, and that the time thus defined is a measure of motion of the particles inside the system in accordance with the spirit of Einstein's light clock referred to above. (See Refs. \cite{Ki}, \cite{Ki2} for these facts.) We remark that the same definition as ours is adopted by Ahluwalia et. al. in Refs. \cite{Ah}, \cite{Ah2}, \cite{AAh}, where our clock is called ``flavor-oscillation clock" and the effect of gravity in quantum mechanical observation is discussed. Their ``characteristic time of flavor-oscillation" (eq. (10) in Ref. \cite{Ah}) is exactly a correspondent to our ``least period of time" which we will define in the following.
\BP

\F
If $t$ is given as such, we can describe the QM motion of the particle by a solution
\beq
\phi(t) = \exp(-itH/\hbar)\phi(0). \label{(3)}
\ene
of the Schr\"odinger equation
\beq
\frac{\hbar}{i}\frac{d}{dt}\phi(t) + H \phi(t) = 0. \label{(2)}
\ene

\MP

\F
If we insert the QM velocity $P/m$ into $u$ in postulates A1 and A2 above, then they become
\MP

\F
A1. $|P/m|^2+|v|^2=c^2$.
\MP

\F
A2. $|P|^2=(m_0)^2c^2$.
\BP

\F
But $P$ is a differential operator and the rest are numerical quantities, so these
conditions are meaningless. To make these two postulates meaningful, we move to a
momentum space by spectral representation or by Fourier transformation as
follows.
\MP

\F
In the one particle case, the particle is free from interaction, and the solution
$\phi(t)=\exp(-itH/\hbar)\phi(0)$ is given by using Fourier transformation ${\cal F}$:
\beq
{\cal F}f(p)=(2\pi \hbar)^{-3/2}\int_{R^3} \exp(-ip\cdot x/\hbar)f(x)dx, \label{(4)}
\ene
where $p \in R^3$, $p\cdot x=p_1x_1+p_2x_2+p_3x_3$,
as follows:
\beq
\phi(t)=\phi(t,x)={\cal F}^{-1}\exp(-itp^2/(2m\hbar)){\cal F}\phi(0). \label{(5)}
\ene
Also the Hamiltonian $H$ is given by
\beq
H=\frac{1}{2m}{\cal F}^{-1}p^2{\cal F}. \label{(6)}
\ene
${\cal F}$ is a unitary operator from ${\cal H}=L^2(R^3)$ onto itself. Here $L^2(R^3)$ is the Hilbert
space of Lebesgue measurable complex-valued functions $f(x)$ on $R^3$ that satisfies
$$
\int_{R^3}|f(x)|^2dx < \infty,
$$
and has inner product and norm:
$$
(f,g)=\int_{R^3}f(x)\overline{g(x)} dx,\quad \Vert f\Vert=\sqrt{(f,f)},
$$
where
$\overline{g(x)}$ is the complex conjugate to a complex number $g(x)$.
\MP

\F
From this we construct a spectral representation of $H$ as follows:
\MP

\F
Let ${\cal F}(\lambda)$ $(\lambda>0)$ be a map from a subspace (exactly speaking, $L^2_{s}(R^3)$ with
$s>1/2$, see Ref. \cite{Ki3}) of $L^2(R^3)$ into
$L^2(S^2)$ ($S^2$ is the unit sphere in $R^3$) defined by
\beq
{\cal F}(\lambda)f(\omega)=(2\lambda)^{1/4}({\cal F}f)(\sqrt{2\lambda}\omega), \label{(7)}
\ene
where $\omega$ is in $S^2$, i.e. $\omega \in R^3$ and $|\omega|=1$.
Then by \eq{(6)} we have
\beq
{\cal F}(\lambda)Hf(\omega)=(\lambda/m){\cal F}(\lambda)f(\omega). \label{(8)}
\ene
Thus $H$ is identified with a multiplication by $\lambda/m$ when transformed by
${\cal F}(\lambda)$ into a spectral representation space
$$
\widehat{{\cal H}} = L^2((0,\infty), L^2(S^2), d\lambda),
$$
where $\{{\cal F}(\lambda)\}_{\lambda>0}$ is extended to a unitary operator from ${\cal H}$ onto $\widehat{\cal H}$ in the following sense
$$
\int_0^\infty\Vert {\cal F}(\lambda)f\Vert_{L^2(S^2)}^2 d\lambda =\Vert f\Vert_{L^2(R^3)}^2.
$$
(For more details, see Ref. \cite{Ki3}
Chapter 5, section 5.1.)
\MP

\F
Summing up we can regard $H$ as a multiplication operator $\lambda/m$ by moving to a
momentum space representation.
\MP

\F
Originally, $H$ is
$$
H = \frac{1}{2m}P^2, \eqno{(1)}
$$
and $P$ is a 3 dimensional QM momentum. Thus formally we have a correspondence
\beq
\lambda \leftrightarrow P^2/2.   \label{(9)}
\ene
Thus if we denote the QM velocity of the particle inside the local system by $u$,
it is
$$
u=P/m
$$
and satisfies a relation
\beq
u^2=P^2/m^2 \leftrightarrow 2\lambda/m^2. \label{(10)}
\ene
Therefore our postulates A1 and A2 above are restated as follows:
\MP

\F
A1. $2\lambda/m^2 + |v|^2=c^2$.
\MP

\F
A2. $2\lambda=(m_0)^2c^2$.
\MP

\F
These are now meaningful, as the relevant quantities are all numeric. These
axioms correspond to a requirement that we think all things in the local system of the one particle, on an energy shell
$H=\lambda/m=(m_0)^2c^2/(2m)$ of the Hamiltonian $H$, whenever considering the observation of the particle.
\MP

\F
The first consequence of these formulation is
\BP

\F
Proposition 2. $m=m_0/\sqrt{1-(v/c)^2} (\ge m_0)$.
\MP

\F
Proof. From A1 and A2 follows
$$
(m_0)^2c^2+m^2|v|^2=m^2c^2,
$$
which yields the proposition. QED
\BP

\F
We define:
\MP

\F
Definition 3. The period $p(v)$ of a local system moving with velocity $v$ relative
to the observer is defined by the relation:
$$
p(v)\lambda/(\hbar m)=2\pi.
$$
Thus
\beq
p(v) = hm/\lambda = 2hm/[(m_0)^2c^2]. \label{(11)}
\ene
This gives a period of the local system with the clock on the energy shell
$H=\lambda/m$:
$$
exp(-itH/\hbar)=exp(-it\lambda/(\hbar m)).
$$
In particular, when $v=0$, the period $p(0)$ takes the minimum value:
$$
p(0)=hm_0/\lambda=2hm_0/[(m_0)^2c^2]=2h/(m_0 c^2).
$$
This we call the least period of time (LPT) of
 the local system. This gives a minimum cycle or period
 proper to the local system. 
This combined with the discussion of Ahluwalia\cite{Ah}
 of rotating gravitational field would give a basis for
 P. Beamish's RBT (Rhythm Based Time)\cite{RBT2000}.

\MP

\F
The general $p(v)$ is related to this by virtue of Proposition 2 as follows:
\BP

\F
Proposition 4.
\beq
p(v)=hm/\lambda=p(0)/\sqrt{1-(v/c)^2} (\ge p(0)). \label{(12)}
\ene
This means that the time $p(v)$ that the clock of a local system,
 moving with velocity $v$ relative to the observer, rounds $1$ cycle
 when it is seen from the observer, is longer than the time $p(0)$
 that the observer's clock rounds $1$ cycle, and the ratio is given
 by $p(v)/p(0)=1/\sqrt{1-(v/c)^2} (\ge 1)$. Thus time, measured
 by our QM clock, of a local system moving with velocity $v$
 relative to the observer becomes slow with the rate
 $\sqrt{1-(v/c)^2}$, which is exactly the same as the rate
 that the special theory of relativity gives. This yields that
 the QM clock obeys the same transformation rule as that for
 classical relativistic clocks like light clock discussed
 in the introduction, and shows that quantum mechanical clock
 is equivalent to the relativistic classical clock.
\BP

These mean also that the space-time measured by using
 QM clock defined as the QM evolution of a local
 system follows the classical relativistic change of coordinates of
 space-time. Thus giving a consistent unification of QM and special
 relativistic CM. 
\BP

As for the validness of the name LPT, we see how it gives the Planck time:
$$
t_P=\sqrt{Gh/c^5} = 1.35125\times 10^{-43}{\rm\ s},
$$
where $G$ is the gravitational constant. In fact, given Planck mass:
$$
m_0=m_P = \sqrt{hc/G} = 5.45604\times 10^{-5} {\rm\ g},
$$
our LPT yields
$$
p(0) = 2h/(\sqrt{hc/G}c^2)
= 2\sqrt{hG/c^5} = 2t_P,
$$
which is 2 times Planck time.
\BP

\vskip12pt

%\pagebreak

\F
{\bf II. $N$-particles case with masses $m_j$ $(j=1,2,\cdots,N)$}
\BP

\F
We consider the case after the local system $L$ of $N$ particles are scattered sufficiently. Then the
system's solution asymptotically behaves as follows as the system's time $t=t_L$ goes to $\infty$ 
(see Ref. \cite{Ki2}):
$$
\exp(-it_LH_L/\hbar)f
\sim \exp(-it_L h_b/\hbar)g_0
\otimes\exp(-it_LH_1/\hbar)g_1\otimes\cdots\otimes \exp(-it_LH_k/\hbar)g_k,
$$
where $h_b=T_b+I_b(x_b,0)$ and $k \ge 1$.
\MP

\F
For explanations of notations, see Ref. \cite{Ki2}. Here it will suffice to note
that the Hamiltonians $H_\ell$ $(\ell=1,2,...,k)$ of each scattered cluster are treated just
as in the case I) above but with using a general theorem on the spectral
representation of self-adjoint operators $H_\ell$. $H_\ell$ are not necessarily free
Hamiltonians and we cannot use the Fourier transformation, but we can use
spectral representation theorem so that each $H_\ell$ is expressed, unitarily equivalently, as $\lambda_\ell/M_\ell$ in
some appropriate Hilbert space, where $M_\ell$ is the mass of the $\ell$-th cluster.
\MP

\F
Then it is done quite analogously to I) to derive the propositions 2 and 4 above
in the present case. Of course these relations depend on $\ell$, and show that the dependence of mass and time of each cluster on the relative velocity to the observer is exactly the same as the special theory of relativity gives as in the case I) above.
\BP

One of the important consequences of these arguments is that the quantum clock is equal to the classical relativistic clock, which has remained unexplained as one of the greatest mysteries in modern physics in spite of the observed fact that they coincide with high precision.
%\BP

%\vskip12pt

%\BP

%\BP

%\BP

%\vskip10pt

\end{document}